\documentclass[pra,aps,twocolumn,showpacs]{revtex4}

\usepackage{graphicx} 
\usepackage{amsmath}
 
\newcommand{\beq}{\begin{equation}}
\newcommand{\enq}{\end{equation}}
\newcommand{\la}{\langle}
\newcommand{\ra}{\rangle}

\begin{document}

\title{Ultracold polarized Fermi gas at intermediate temperatures}
\author{Jani-Petri Martikainen}
\email{Jani-Petri.Martikainen@helsinki.fi}
\affiliation{Department of Physical Sciences, University of Helsinki, 
PO Box 64, 00014 University of Helsinki,  Finland}
\date{\today}
\begin{abstract}
We consider  non-zero temperature properties of the polarized two-component Fermi gas. 
We point out that stable polarized
paired states which are more stable than their phase separated
counterparts with unpolarized superfluid region can exist below the critical temperature.
We also solve the system behavior in a trap using the local density approximation
and find gradually increasing polarization in the center of the system
as the temperature is increased. However, in the strongly interacting region
the central polarization increases most rapidly close to the mean-field
critical temperature, which is known to be substantially higher than
the critical temperature for superfluidity. This indicates that
most of the phase separation occurs in the fluctuation region
prior to superfluidity and that the polarization in the actual superfluid is
modest.
\end{abstract}
\pacs{03.75.-b, 32.80.Pj, 03.65.-w}  
\maketitle

\section{Introduction}
\label{sec:intro}

Since their experimental realization the properties of the strongly interacting superfluid Fermi 
gases~\cite{Ohara2002b,Jochim2003b,Greiner2003a,Regal2004a,Zwierlein2004a,Bartenstein2004b,Kinast2004a,Chin2004a,Kinast2005a,Zwierlein2005a} 
have been studied ever more actively.
In particular the recent experiments~\cite{Zwierlein2006a,Partridge2006a,Zwierlein2006b,Partridge2006b,Zwierlein2006c}
in a {\it polarized} two-component Fermi system of trapped ultra cold gases
have probed the properties of fermion pairing in a novel settings.
Some important theoretical contributions~\cite{Mizushima2005b,Sheehy2006a} appeared
before the breakthrough experiments, but the experiments motivated a series of theoretical 
works~\cite{Kinnunen2006a,Pieri2006a,Yi2006a,Chevy2006a,DeSilva2006a,Haque2006a,Pao2006a} 
which have explained various features of the experiments with non-rotating gases 
in the low temperature regime. 

Since the BCS-phase cannot support
polarization at zero temperature, one typically expects, in a harmonic trap, phase
separation  into an unpolarized superfluid core, surrounded by the polarized normal 
Fermi gases. The transition between the phases is sharp in the local density
approximation (LDA) and local minimization of the grand potential rules 
out the possibility of the local maxima of the free energy,
known as the Sarma~\cite{Sarma1963a} or breach-pair (BP)~\cite{Forbes2005a} state.
LDA approach can, however, be modified to include a possibility for a 
BP phase if one minimizes the total (global) energy of the system~\cite{Jensen2006a}.
Such modification can in principle predict a ring of BP phase in the boundary region
where BCS phase is transformed into the normal phase, and for large polarization, BP phase is
also found in the center of the trap. In this paper
we will assume that these boundary effects are small and can be ignored. For
a system in which the number of fermions is very large and the trap does not have
a too high aspect ratio, this is expected to be
a good approximation although it is quite likely that these boundary terms
are important in defining the largest  possible polarization that can still
support a superfluid.

While the above described works consider mainly
zero temperature, we now concentrate on finite temperature
effects. In a work fairly closely related to ours Machida {\it et al.}~\cite{Machida2006a} 
considered the finite temperature effects by solving the mean-field Bogoliubov-de Gennes equations
numerically in a cylinder with a harmonic trap in the radial direction. Qualitative
picture they find is in many places similar to ours, although the simple local density approximation we apply
in a trap is insufficient to properly describe the regions they call ``FFLO'' phase.
Also  a very recent work by Chien {\it et al.}~\cite{Chien2006a} investigated the 
superfluid properties of the homogeneous polarized Fermi gas across the BEC-BCS cross-over
also at non-zero temperatures. At intermediate temperature regime they found 
that stable polarized Fermi superfluids are  possible. Various aspects of finite temperature
effects in BEC-BCS cross-over were also recently discussed by Yi and Duan~\cite{Yi2006b}.

The purpose of this paper
is to point out that polarized BCS states, when they exist, also have a lower energy than
the corresponding phase separated state with an (finite temperature) unpolarized
paired part and polarized normal part. Furthermore, using LDA we find the 
state of the system in a trap as a function of temperature and find 
that the system becomes gradually  polarized 
in the center of the trap. Since this occurs also at temperatures still well below the critical 
temperature we expect the mean-field theory employed here to provide at least
qualitatively valid description of the system. However, most rapid increase of the 
central polarization occurs in the fluctuation region above the superfluid transition
temperature. This indicates that most of the phase separation in  the trapped system 
would occur prior to the onset of superfluidity.

This paper is organized as follows. In Sec.~\ref{sec:BCS}
we discuss the uniform BCS phase and compute how 
highly polarized such a state could be.
In Sec.~\ref{sec:Separated} we proceed to compare the 
energies of the polarized BCS phase with the mixed phase of unpolarized
BCS phase and a polarized normal gas.
In Sec.~\ref{sec:LDA} we investigate the intermediate temperature properties
in a trapped gas using the local density approximation.
We end with some concluding remarks in Sec.~\ref{sec:Conclusions}.

\section{Polarized uniform density BCS state}
\label{sec:BCS}
The grand potential corresponding to the standard mean-field
BCS theory of interacting two component Fermi can be computed
from the mean-field Hamiltonian
\begin{eqnarray}
H&=&\sum_{{\bf k},\sigma}\left[\epsilon_{\bf k}-\mu_\sigma\right]
{\hat\psi}_{{\bf k},\sigma}^\dagger{\hat\psi}_{{\bf k},\sigma}
\\
&+&\Delta \sum_{{\bf k}} \left({\hat\psi}_{{\bf k},2}^\dagger
{\hat\psi}_{-{\bf k},1}^\dagger
+{\hat\psi}_{{\bf k},1}{\hat\psi}_{-{\bf k},2}\right)
-\frac{\Delta^2}{g},\nonumber 
\end{eqnarray}
where ${\hat\psi}_{{\bf k},\sigma}$ is the (fermionic) annihilation
operator for species $\sigma$ and
the coupling strength $g=4\pi\hbar^2 a/m$ is expressed in terms of
the scattering length $a$. Furthermore, 
$\Delta=g\la {\hat\psi}_{1}({\bf r}){\hat\psi}_{2}({\bf r})\ra$
is the order parameter/gap function, 
which we take to be independent of position, $\epsilon_k=\hbar^2k^2/2m$
is the free dispersion,
and $\mu_\sigma$ is the chemical potential for the $\sigma$-component.
Since we are going to apply this Hamiltonian also in the strongly interacting
regime, some phenomenological model building is clearly involved. 
In particular, the existence of a non-zero
gap function should not be taken as an indicator of superfluidity. 
Superfluidity would imply phase stiffness and this typically occurs
at a lower temperature~\cite{Milstein2002a,Perali2004a} than the mean-field critical temperature predicted
by the above Hamiltonian.

The assumption of a constant gap function is in this context
quite reasonable, but it does rule out the possibility of the 
FFLO-phases~\cite{Fulde1964a,Larkin1964a}, which could 
in principle exist in some narrow range of parameters~\cite{Sheehy2006a}.
Diagonalizing the above Hamiltonian we can  simply calculate
the grand canonical free energy
\beq
\label{eq:fe_1}
\Omega\left(\mu_1,\mu_2,\Delta\right)=-k_BT\log[Tr(\exp(-\beta H))],
\enq
as a function of temperature $T$~\cite{Yi2006a}. Due to the use
of the contact interaction the resulting expression is ultraviolet divergent
and this divergence is removed by the usual renormalization
\beq
\frac{\Delta^2}{g}\rightarrow 
\frac{\Delta^2}{g}-\frac{1}{V}\sum_{{\bf k}}\frac{\Delta^2}{2\epsilon_{\bf k}}
\enq
where $V$ is the quantization volume which drops out from the final physical results.
The gap equation can be derived from the grand potential as condition for
the extrema $\partial\Omega/\partial\Delta=0$, while the number equation
appears from $n_\sigma=-\partial\Omega/\partial\mu_\sigma$, where $n_\sigma$ is the component density. 
Alternatively, the latter condition
can be enforced as an extremal point of the Helmholtz free energy $F=\Omega+\mu_1n_1+\mu_2n_2$,
i.e. $\partial F/\partial\mu_\sigma=0$.
In this section we choose the units using the ideal Fermi gas of density ${\bar n}$
as a benchmark and ${\bar n}$ is taken to be an average of the component
densities. This means that  the unit of length is given by
$l=1/k_F=(6\pi^2 {\bar n})^{-1/3}$, while the unit of energy is $\epsilon_F=\hbar^2k_F^2/2m$.

The polarized BCS solution corresponds to the minimum of the 
grand canonical free energy, but does not exist at zero temperature.
At zero temperature only stationary points of the free energy corresponding to
the polarized solutions and spatially constant gap 
are the normal state and the BP state. However, as
the temperature is increased stable polarized BCS solutions can appear.

\begin{figure} 
\includegraphics[width=0.90\columnwidth]{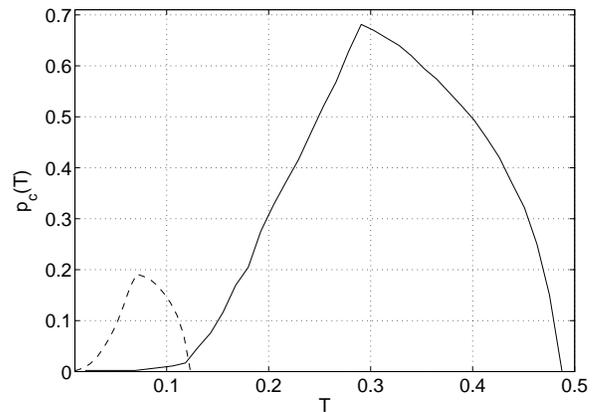} 
\caption[Fig1]{The highest value of polarization that the uniform BCS state can support
as a function of temperature (in units of Fermi temperature) when $k_Fa=-100$ (solid line) and $k_Fa=-1$ (dashed line).
We have set the maximum polarization equal to zero in the region above the critical temperature.
}
\label{fig:critpol}
\end{figure}

In Fig.~\ref{fig:critpol} we show the largest possible polarization $p_c(T)=(n_1-n_2)/(n_1+n_2)$ 
with a large and moderate coupling strength.
It  is clear that BCS solutions corresponding to the minimum of the grand potential with substantial
polarizations do exist in the strongly interacting regime as the temperature increases. 
For very weak interactions the polarization supported becomes vanishingly small, but
the maximum polarization of the BCS state can be as high as $p_c(T)\approx 0.68$ 
in the strongly interacting regime with $k_Fa=-100$. If the polarization is higher
than this the system will never enter a uniform superfluid region as the temperature is lowered
towards $T=0$.
Curiously, this critical {\it density} asymmetry is close to 
the experimentally observed critical {\it atom number} asymmetry in a trap~\cite{Zwierlein2006c}.

Naturally close to critical temperature of the mean-field theory 
the approach used here becomes unreliable, but states of considerable polarization do exist even well below
the critical temperature. The maximum polarization supported by the BCS state is predicted to peak
in the intermediate temperature regime. This is  expected since the critical temperature is
maximized in an unpolarized system and therefore just below the critical temperature only small
polarization is possible.

\section{Phase separated state vs. polarized BCS state}
\label{sec:Separated}
Let us now turn to the possibility of phase separation. 
At  zero temperature the phase separated state is clearly favored
since the polarized BCS solution does not exist. However, as we saw
in the previous section at non-zero temperature the polarized BCS solutions
do exist and it is not clear if these polarized BCS states have a lower
energy than the phase separated state corresponding to the same number of atoms.

In particular, we have in mind a comparison with an unpolarized BCS phase occupying a
volume fraction $1-x$ ($0\le x\le 1$) next to a polarized normal
phase in a volume fraction $x$~\cite{Bedaque2003a}. If the 
unpolarized BCS phase has a density $n_{BCS}$ the normal
state will have fermion densities
\beq
n_{\sigma,n}=\frac{n_\sigma-(1-x)n_{BCS}}{x}
\enq
if the average density for each component, averaged over the whole the system, 
is $n_\sigma$. The problem is then to find the values of 
$x$  and $n_{BCS}$ which minimize the Helmholtz free energy
$F(x,n_{BCS})$ for given set of parameters, such as
coupling strength, polarization, and temperature. 
When the interface energy is ignored the Helmholtz 
free energy is simply a sum of two contributions,
the polarized normal gas and unpolarized BCS phase.

\begin{figure}
\includegraphics[width=0.90\columnwidth]{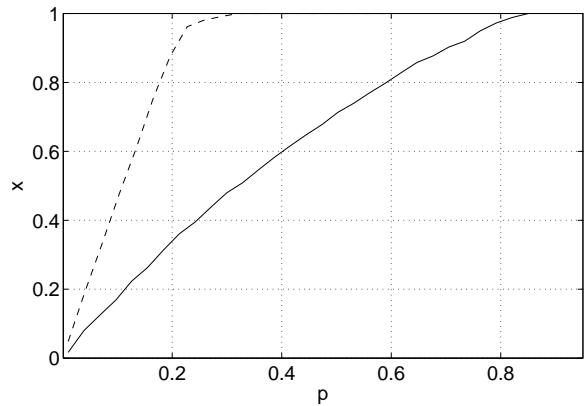} 
\caption[Fig2]{The optimal volume fraction $x$ for the mixed state of the
unpolarized BCS phase and normal polarized gas at $T=0$ as a function
of polarization for $k_Fa=-1$ (dashed) $k_Fa=-10$ (solid).
}
\label{fig:phasesep}
\end{figure}

In Fig.~\ref{fig:phasesep} we show how the optimum normal state
volume fraction behaves as a function of polarization in a phase separated state at zero temperature. 
Quite sensibly,
for weaker interactions the normal part increases faster with polarization
while in the strongly interacting regime the paired part
persists longer. Our result for weaker interaction appears to be consistent
with the earlier result by Bedaque {\it et al.}~\cite{Bedaque2003a}.
As we increase the temperature the normal part will 
become bigger and eventually at the critical temperature we have a phase transition
into a pure normal state. 

Since our approach does not include surface energy contribution
from the boundary between superfluid and normal regions, the zero temperature transition to a fully 
normal state as polarization increases is not sharp. Nevertheless, the volume fraction
of the paired region becomes so close to zero that it drops below our numerical resolution.
If one were to include the surface terms~\cite{DeSilva2006b}, one would expect a sharp boundary
as well as a shift downwards from the ``critical'' polarization around $p\approx 0.85$
for the stronger interactions.

\begin{figure}
\includegraphics[width=0.90\columnwidth]{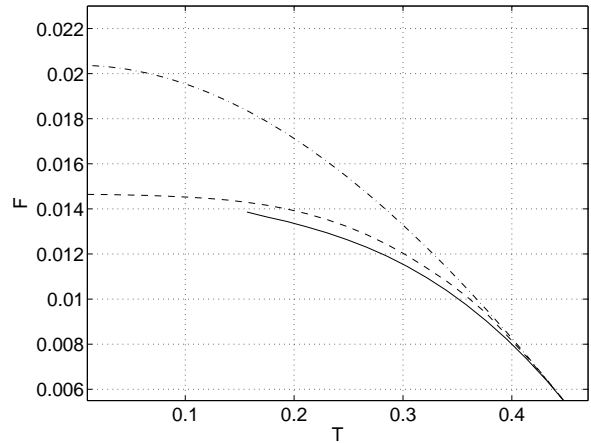} 
\caption[Fig3]{The free energy of the uniform polarized state (solid line),
phase separated state (dashed line), and the polarized normal state (dash-dotted line)
as a function of temperature (in units of Fermi temperature). 
We choose the polarization $p=0.1$ and coupling $k_Fa=-20$.
The line for the uniform polarized state begins from the temperature where the 
uniform polarized phase is first possible.
}
\label{fig:energy_compare}
\end{figure}

In Fig.~\ref{fig:energy_compare} we show the minimum attainable Helmholtz free energy
as a function of temperature
for the phase separated state, together with the energies of the polarized BCS state, and the
pure normal state with fixed polarization of $0.1$ in the strongly interacting regime.
While the energy differences are not large, it is nevertheless clear
that whenever the (stable) polarized paired state exists, it is  more stable
than the phase separated state. Furthermore, the energy of the phase separated state
is  an underestimate due to the absence of the surface energy contributions.
The mean-field theory used here ignores some fluctuations contributions to the
free energy, but we believe these are not going to change the outcome of the 
energy comparison since these contributions would have to be included in both competing states.
This outcome is made all the more likely by the absence of surface energy contributions in the
phase separated state.

\section{Local density approximation at $T\neq 0$}
\label{sec:LDA}
At zero temperature, 
for an ideal Fermi gas we find in the local density approximation
a component density
\beq
n_\sigma({\bf r})=\frac{1}{6\pi^2}\left[\frac{2m}{\hbar^2}
\left(\mu_{\sigma}-V_T({\bf r})\right)\right]^{3/2},
\enq
where the external trapping potential is given by $V_T({\bf r})=m\omega^2r^2/2$ and
the component chemical potential $\mu_\sigma$ is related to the
particle number $N_\sigma$ through $\mu_\sigma=\hbar\omega (6N_\sigma)^{1/3}$.
The density vanishes outside the Thomas-Fermi radius $R_{TF,\sigma}=\sqrt{2\mu_\sigma/m\omega^2}$.
The polarization $p_N=(n_1-n_2)/(n_1+n_2)$ will then be 
\beq
p_N(r)=
\frac{(\mu_1-V_T(r))^{3/2}-(\mu_2-V_T(r))^{3/2}}{
(\mu_1-V_T(r))^{3/2}+(\mu_2-V_T(r))^{3/2}}
\enq
inside the Thomas-Fermi radius of the minority component.
This implies that for an ideal gas the polarization will increase from
\beq
p_N(r=0)=\frac{1-\sqrt{N_2/N_1}}{1+\sqrt{N_2/N_1}}
\enq
in the center of the trap to the maximum value of $p_N(r>R_{TF,2})=1$
beyond the Thomas-Fermi radius of the minority component.

On the other hand at zero temperature superfluid one expects strong
phase separation due to the absence of the polarized
superfluid state corresponding to the minimum of the grand potential.
In that case, if we ignore the boundary effects, we would find
vanishing polarization $p_S(r)$ in the center of the cloud, which then
abruptly increases close to the trap edge, eventually 
becoming $p_S(r>R_{TF,2})=1$ 
beyond the Thomas-Fermi radius of the minority component~\cite{Haque2006a}.

As we have found, the polarized paired state can exist
at non-zero temperature and furthermore when it exists
it has a lower energy than the corresponding phase separated
state with non-polarized superfluid mixed with a normal
Fermi gas. This then raises the question of what the
polarization $p(r)$ will be in a trap 
at intermediate temperatures. Finite temperature effects 
will round off the density distribution and make the
LDA analysis less accurate, but we expect it 
to be qualitatively valid. 
Also, since the mean-field theory
of the strongly interacting Fermi gas becomes more unreliable at
higher temperatures, it appears fruitless to aim
towards higher precision within the framework of the mean-field theory.
 
Since the energy of the polarized BCS state increases with
polarization it will be energetically
more favorable to have a large number of atoms in a region
of fairly small polarization, rather than forming, for example,
a small core of highly polarized superfluid surrounded by the
normal gas or perhaps unpolarized BCS phase.
This implies, quite sensibly, that $p(r)$ wíll be 
something in between the normal state polarization $p_N(r)$
and the $T=0$ phase separated superfluid polarization $p_S(r)$.

The normal state polarization increases smoothly up to $1$
and as we have found, the BCS state can be polarized  if the
(local) polarization is below the critical value $p_c(T)$. 
Also one expects that the extent of the (possibly) polarized superfluid core
becomes smaller as the temperature is increased and eventually disappears
at a critical temperature where polarized BCS states
can no longer be found.
Outside the paired core one expects a polarized  normal Fermi gas. 
These qualitative arguments
are quite well confirmed by solving the state of the system using LDA analysis.  This implies
using the local chemical potentials $\mu_{eff,\sigma}=\mu_\sigma-V_T(r)$
and finding $\mu_\sigma$ in such a way that the total particle numbers have the desired values.
At non-zero temperatures the contributions from $\mu_{eff,\sigma}<0$
do not vanish identically and the edge of the cloud is rounded somewhat by this effect.

\begin{figure}
\includegraphics[width=0.98\columnwidth]{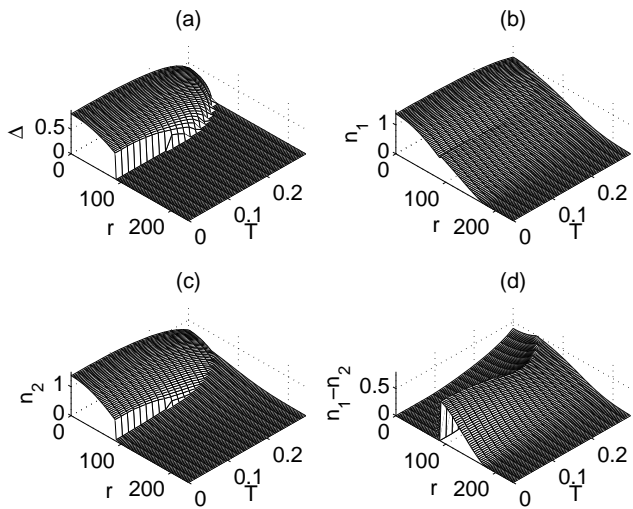} 
\caption[Fig4]{(a) The pairing gap, (b)-(c) component densities, and
(d) the density difference as a function of position and temperature.
The result is calculated in the strongly interacting regime
with coupling strength $k_Fa=-20$ and number asymmetry $(N_1-N_2)/(N_1+N_2)=0.5$. 
The unit of energy is the 
Fermi energy $\epsilon_F=\hbar k_F^2/2m$ of the ideal Fermi gas of 
$^{6}{\rm Li}$ atoms with $(N_1+N_2)/2=10^5$ atoms in a harmonic trap with 
$2\pi (\omega_x\omega_y\omega_z)^{1/3}=600\, {\rm Hz}$.
Unit of length is $1/k_F$ and densities are scaled with the factor $6\pi^2$ so that
ideal gas would have a density of $n_{ideal}=1$ in the center.
}
\label{fig:LDAsolution}
\end{figure}

In Fig.~\ref{fig:LDAsolution} we show a typical result for the densities, the gap, as well as the 
density difference as a function of position and temperature in the strongly 
interacting regime with a number asymmetry $(N_1-N_2)/(N_1+N_2)=0.5$. 
The behavior
of the gap function is qualitatively similar even if the paired region were
polarized. The spatial extent of the paired core first, somewhat surprisingly, increases slightly as the 
temperature increases, but is then reduced as one approaches the critical temperature.
This effect is however so small, that it would be difficult to observe and 
could be an artifact of the LDA. At zero temperature LDA analysis gives rise
to abrupt changes in densities (especially for the minority component)
and also the gap drops suddenly to zero. Increasing temperature eventually 
removes all these discontinuities. As the temperature is  lowered
below the critical temperature the minority component develops a clear bimodal density
distribution.

The polarization as a function of position naturally becomes different
as the temperature increases due to the 
appearance of polarization in the paired core. The polarization
in the center of the cloud increases gradually with increasing temperature, but 
this build-up becomes more abrupt if the number asymmetry increases.
If fluctuation effects were to be included, the critical temperature for the
superfluidity could become so low~\cite{Chien2006a} (in the strongly interaction region)
that the actual superfluid region would have
only a small polarization. The build-up of polarization and removal
of phase separation with increasing
temperature would therefore mostly occur in the ``pseudo-gap'' region. 
For more weakly interacting system this distinction disappears, but
then also the polarization that the BCS state can carry drops and 
the superfluid region will again have only a small polarization.

Qualitatively the behavior predicted by the LDA does not change strongly 
when the number asymmetry is increased to $0.8$. Interestingly, it turns out that 
the position dependence of the local polarization $\left[n_1(r)-n_2(2)\right]/\left[n_1(r)+n_2(r)\right]$
becomes weaker as the temperature increases. This suggests that
approximating the system as uniform, might not always 
be too drastic approximation, especially in the neighborhood of the temperature when pairing effects
begin.

\begin{figure}
\vspace{1cm}
\includegraphics[width=0.98\columnwidth]{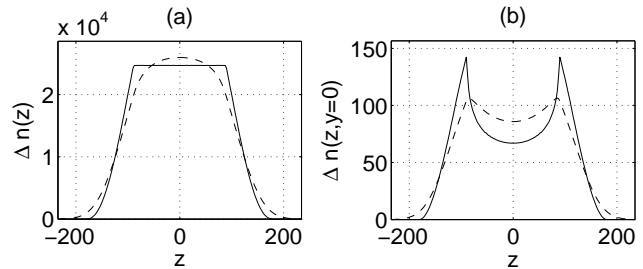} 
\caption[Fig5]{(a) The doubly integrated density difference  
$\Delta n(z)=\int dxdy \left[n_1({\bf r})-n_2({\bf r})\right]$ at 
$T=0$ (solid) and $T=0.2$ (dashed). (b) The column integrated density difference  
$\Delta n(z,y=0)=\int dx \left[n_1(x,y=0,z)-n_2(x,y=0,z)\right]$ at 
$T=0$ (solid) and $T=0.2$ (dashed).
The result is calculated with same parameters and units as in
Fig.~\ref{fig:LDAsolution}.
}
\label{fig:Integrated_Dens}
\end{figure}

Since it is experimentally easier to measure integrated densities
such as the doubly integrated $n_\sigma(z)=\int dxdy n_\sigma(x,y,z)$ or the
column integrated $n_\sigma(z,y=0)=\int dx n_\sigma(x,y=0,z)$ we show in 
Fig.~\ref{fig:Integrated_Dens} the integrated density differences
for the polarized and unpolarized superfluid systems at zero as well as
non-zero temperature. As can be seen the signals are not qualitatively
very different. The polarized state has somewhat higher central
density difference and finite temperature effects make the profiles of the polarized
system more smooth.

\section{Summary and conclusions}
\label{sec:Conclusions}
In this paper we have employed the mean-field theory at finite temperatures
to discuss the behavior of the interacting Fermi gases with unequal numbers
of fermions in two components. We found that stable polarized 
paired states can exist at intermediate temperatures and that cross-over
to a polarized paired state can leave a moderate signature
on the integrated density distribution of the Fermi gas.
The LDA approach used here predicts a gradual appearance of the 
polarized superfluid as the temperature is increased. 

The accuracy of the mean-field theory in the strongly interacting regime is naturally questionable
especially at non-zero temperatures and is expected to give rise,
for example, to a substantial overestimate of the critical temperature for superfluidity~\cite{Milstein2002a}.
Therefore, the critical temperature found here provides an estimate on the temperature when
pairing effects appear, but not necessarily on the superfluid transition temperature
when phase coherence sets in~\cite{Perali2004a}. The results presented here in the strongly interacting region
are qualitatively similar in the weakly interacting regime where the fluctuations are less strong.
The absolute magnitudes of the possible polarizations, the superfluid gap, and the
critical temperature are then, however, substantially smaller.

At zero temperature it is known from the Monte-Carlo calculations~\cite{Carlson2005a}
that the mean-field theory overestimates the gap substantially.  In order to provide more accurate quantitative predictions
it would therefore be important to understand the intermediate temperature
properties of strongly interacting Fermi gases better. This is all the more 
important due to the very recent MIT experiment~\cite{Zwierlein2006c} which observed 
prominent interaction effects close to and somewhat above 
the (superfluid) transition temperature and also pointed
out the need of incorporating the interaction effects between the 
normal and superfluid regions into a quantitatively accurate theory.
  
\begin{acknowledgments}
This work was supported by Academy of Finland
(Academy project number 207083).
\end{acknowledgments}

\bibliographystyle{apsrev}

\bibliography{bibli}

\end{document}